\documentclass[11pt]{article} 
\usepackage{epsfig,amssymb}
\usepackage{amsmath}
\usepackage{natbib}
\usepackage{graphics}
\usepackage{graphicx}
\usepackage{dcolumn}
\usepackage{bm}
\usepackage{epsfig}

\begin{document}

\begin{center}

{{\bf Opportunities for Probing Fundamental Gravity with Solar System Experiments}\\
A Science White Paper\footnote{Contact information: 
Dr. Slava G. Turyshev, Jet Propulsion Laboratory, 4800 Oak Grove Drive, Pasadena, CA 91109, phone: (818) 393-2600, email:  turyshev@jpl.nasa.gov and 
Prof. Thomas W. Murphy, Jr., University of California, San Diego, 9500 Gilman Drive, La Jolla, CA 92093, phone:  (858) 534-1844, email:  tmurphy@physics.ucsd.edu} submitted to \\
the {\it Cosmology and Fundamental Physics} Science Frontier Panel of Astro2010
}


\normalsize
\bigskip 

Eric G. Adelberger$^a$, James Battat$^b$, Douglas Currie$^c$, William M. Folkner$^d$, Jens Gundlach$^a$, Stephen M. Merkowitz$^e$, Thomas W. Murphy, Jr.$^f$, Kenneth L. Nordtvedt$^g$, \\
Robert D. Reasenberg$^h$, Irwin I. Shapiro$^h$, Michael Shao$^d$, Christopher W. Stubbs$^i$,\\ 
Massimo Tinto$^d$, Slava G. Turyshev$^d$, James G. Williams$^d$, and Nan Yu$^d$

\normalsize
\bigskip 

{\it $^a$Center for Experimental Nuclear Physics and Astrophysics,\\
University of Washington, Seattle, WA 98195-4290, USA}\\[3pt]

{\it $^b$Department of Physics, Massachusetts Institute of Technology, Cambridge, MA 02139, USA}\\[3pt]

{\it $^c$Physics Department, University of Maryland, College Park, MD 20742, USA}\\[3pt]

{\it $^d$Jet Propulsion Laboratory, California Institute of Technology,\\
4800 Oak Grove Drive, Pasadena, CA 91109-0899, USA}\\[3pt]

{\it $^e$NASA Goddard Space Flight Center, Greenbelt MD 20771, USA}\\[3pt]

{\it $^f$Center for Astrophysics and Space Sciences, University of California, San Diego,\\
9500 Gilman Drive, La Jolla, CA 92093-0424, USA}\\[3pt]

{\it $^g$Northwest Analysis, 118 Sourdough Ridge Road, Bozeman MT 59715, USA}\\[3pt]

{\it $^h$Harvard-Smithsonian Center for Astrophysics, 60 Garden Street, Cambridge, MA 02138}\\[3pt]

{\it $^i$Department of Physics, Harvard University, Cambridge, MA 02138, USA} 

\end{center}


%
%
%
%
%

\begin{abstract}  
The recent discovery of ``dark energy'' has challenged Einstein's general theory of relativity as a complete model for our macroscopic universe. From a theoretical view, the challenge is even stronger: general relativity clearly does not extend to the very small, where quantum mechanics holds sway. Fundamental physics models thus require some major revisions. We must explore deeper to both constrain and inspire this needed new physics. In the realm of the solar-system, we can effectively probe for small deviations from the predictions of general relativity: Technology now offers a wide range of opportunities to pursue experiments with accuracies orders of magnitude better than yet achieved.  We describe both the relevant theoretical backgrounds and the opportunities for far more accurate solar system experiments.
\end{abstract}

\section{Challenges in understanding the fundamental nature of gravity}

Today physics stands at the threshold of major discoveries.  Growing observational evidence points to the need for new physics.  Efforts to discover new fundamental symmetries, investigations of the limits of established symmetries, tests of the general theory of relativity, searches for gravitational waves, and attempts to understand the nature of dark matter were among the topics at the focus of scientific research at the end of the last century.  These efforts intensified with the unexpected discovery of the accelerated expansion of the universe (i.e., ``dark energy'') made in the late 1990s, triggering many new activities aimed at answering important questions related to the most fundamental laws of Nature.

The fundamental physical laws of Nature are currently described by the Standard Model and Einstein's general theory of relativity.  However, there are important reasons to question the validity of this description. Despite the beauty and simplicity of general relativity and the success of the Standard Model, our present understanding of the fundamental laws of physics has several shortcomings. In particular, if gravity is to be quantized, general relativity will have to be modified; however, the search for a realistic theory of quantum gravity remains a challenge. This continued inability to merge gravity with quantum mechanics together with the challenges posed by the discovery of dark energy indicates that the pure tensor gravity of general relativity needs modification or augmentation. It is believed that new physics is needed to resolve this issue.

Theoretical models of the kinds of new physics that can solve the problems above typically involve new physical interactions, some of which could manifest themselves as violations of the equivalence principle, variation of fundamental constants, modification of the inverse square law of gravity at various distances, Lorenz-symmetry breaking, large-scale gravitational phenomena, and introduce corrections to the current model of spacetime around massive bodies. Each of these manifestations offers an opportunity for experiment and could lead to a major discovery.

On larger scales, the current model of the universe includes critical assumptions, such as an inflationary era in primordial times, and peculiar settings, such as the fine tuning in the hierarchy problem, that call for a deeper theoretical framework.  The very serious vacuum and/or dark energy problems and the related possible cosmological phase transitions also lead researchers to areas beyond general relativity and the Standard Model. Thus, the discovered baryon acoustic oscillations, and the supernovae and CMB data provide important constraints for relevant models and scenarios. Therefore, just as the investigations of dark energy have far-reaching implications for other fields of physics, advances and discoveries made in fundamental physics may point the way to understanding the nature of dark energy.  For instance, such a pointer could come from observing any evidence of a failure of general relativity on any of the accessible scales. 

The Big Question is: {\it Could it be that our description of the gravitational interaction at the relevant scales is not adequate and stands at the root of all or some of the problems above?} Should we consider modifying or extending our theory of gravity and if so, would this help answer the cosmological and astrophysical riddles?

For fundamental physics, our solar system is a laboratory that offers many opportunities to improve tests of relativistic gravity. For strong-gravity regimes, we may use external laboratories in the form of binary pulsars and inspiral mergers. Recent technological advances pertinent to exploration of gravity, together with the availability of quiescent space environments, put us in a position to address some of the pivotal questions of modern science \citep{Turyshev-etal-2007}.

In this white paper we discuss opportunities for major advances in our understanding of fundamental physics that could be realized in the next decade by addressing the important and challenging questions that physics and astronomy face today.

\section{Search for a new theory of gravity and cosmology}

The emphasis in modern gravitational research is on the fundamental questions at the intersection between particle physics and cosmology---including quantum gravity and the very early universe. This work originated ideas on large extra dimensions, large-distance modification of gravity and brane inflation in string theory, all leading to experimentally-testable explanations for the quantum stability of the weak interaction scale. Recent work in various extensions of gravitational models, including brane-world models, and also efforts to modify models of gravity on large scales motivate new searches for experimental signatures of small deviations from general relativity on various scales, including distances in the solar system (see, e.g., \cite{Turyshev-2009}). 

\subsection{Is the Equivalence Principle exact?}

Einstein's Equivalence Principle (EP) lies at the foundation of his general theory of relativity; testing this fundamental assumption with the highest possible sensitivity is clearly important, particularly since it may be expected that the EP will not hold in quantum theories of gravity. 

In its {\it weak form} (the WEP), the Principle states that the gravitational properties of primarily strong and electro-weak interactions obey the EP. In this case the relevant test-body differences relate to their fractional nuclear- and atomic- binding energy differences. 

General relativity and other metric theories of gravity assume that the WEP is exact.  However, many gravitational theories predict observable violations of the EP at various fractional levels of accuracy ranging from $10^{-13}$ down to $10^{-16}$ \citep{Damour-etal-2002}.  For instance, extensions of the Standard Model \citep{Kostelecky-Potting-1995} that contain new macroscopic-range quantum fields predict quantum exchange forces that generically violate the WEP because they couple to generalized ``charges'' rather than to mass/energy alone, as in general relativity.  Therefore, even a confirmation that the WEP is not violated at some level will be exceptionally valuable, placing useful constraints on the range of possibilities in the development of a unified physical theory.

The most accurate results in tests of the composition-independence of acceleration rates of various masses toward the Earth were reported by ground-based laboratories \citep{Adelberger-etal-2003}. A recent experiment measured the fractional differential acceleration between Be and Ti test bodies at the level of $\Delta a/a=(0.3\pm1.8)\times10^{-13}$ \citep{Schlamminger-etal-2008}. Technology now offers a wide range of opportunities to pursue WEP tests with accuracies spanning $10^{-15}-10^{-18}$. To achieve these dramatic gains, new experiments could rely on differential accelerometers, optical metrology, atom interferometers, and/or cryogenically controlled test masses (for a review, see \citep{Turyshev-2008}).

In its {\it strong form} (the SEP) the EP is extended to cover the gravitational properties resulting from gravitational energy itself, thus involving an assumption about the non-linear property of gravitation. In the SEP case, the relevant test body differences are the fractional contributions to their masses by gravitational self-energy. Because of the extreme weakness of gravity, a test of the SEP requires bodies with astronomical sizes.  Although general relativity assumes that the SEP is exact, many modern theories of gravity typically violate the SEP by including new fields of matter, notably scalar fields \citep{Damour-Nordtvedt-1993,Damour-Polyakov-1994}. 

Currently, the Earth-Moon-Sun system provides the best solar system arena for testing the SEP. Lunar laser ranging (LLR) experiments involve reflecting laser beams off retroreflector arrays placed on the moon in 1970s (see, e.g., \citep{Williams-etal-2009}). Recent solutions using LLR data, in combination with laboratory experiments on the WEP, yield an SEP test of $(-1.8 \pm 1.9)\times 10^{-13}$ with the SEP violation parameter $\eta$ found to be $\eta = (4.0 \pm 4.3) \times  10^{-4}$ \citep{Turyshev-Williams-2007}.  Millimeter precision ranges to the moon \citep{Murphy-etal-2008} should translate into order-of-magnitude accuracy gains in these tests. Further improvements may be achieved via new laser ranging instruments deployed on the moon or other planetary surfaces.

On larger scales, other techniques could offer important contributions. Thus, observations of binary pulsars have been very fruitful \citep{Kramer-etal-2007}, providing convincing evidence for the radiation reaction due to emission of gravity waves. The recent discovery of a double-pulsar system opens possibilities for a whole new category of tests of general relativity. An excellent SEP test could come from timing radio signals from an isolated black-hole / pulsar binary system, as the test bodies would have a very large self-energy asymmetry. An SEP violation would show up through production of dipole (as opposed to quadrupole) gravitational radiation.

\subsection{Do the Fundamental Constants of Nature vary with space and time?}

The possibility that fundamental physical parameters may vary with space and time has been revisited with the advent of models unifying the forces of nature based on the symmetry properties of possible extra dimensions, such as Kaluza-Klein-inspired theories, Brans-Dicke theory, and supersymmetry models.  Modern gravitational models that attempt to ``complete'' general relativity at very short distances (embedding it in a more powerful theory capable of addressing phenomena on that scale) or ``extend'' it on very large distances ($\sim10^{28}$ cm) typically include cosmologically evolving scalar fields that lead to variability of the fundamental constants.  Furthermore, it has been hypothesized that a variation of the cosmological scale factor with epoch could lead to temporal or spatial variation of physical constants, specifically the gravitational constant, $G$, the fine-structure constant, $\alpha$, and the electron-proton mass ratio, $m_e/m_p$.

If the values of fundamental constants vary from place to place, they might also be expected to evolve in time. The anticipated space-time variability motivates measurements of the spatial gradients as well as temporal changes of the fundamental physical parameters; besides, measuring the former usually probes theoretical possibilities more strongly.  It is clear that, in the search for the next fundamental theory of gravity, opportunities to push the frontiers in both directions are desirable. Constraints on the variation of fundamental constants can be derived from a number of gravitational measurements (i.e., tests of the EP), the motion of the planets in the solar system, plus stellar and galactic evolution. They are based on the comparison of two time scales, the first (gravitational time) dictated by gravity (ephemerides, stellar ages, etc.), and the second (atomic time) determined by a non-gravitational system (e.g. atomic clocks). For instance, planetary and spacecraft ranging, neutron star binary observations, paleontological and primordial nucleosynthesis data allow one to constrain the relative variation of $G$ \citep{Uzan-2003}.

Lunar and planetary ranging measurements currently lead the search for very small spatial or temporal gradients in the value of G.  Recent analysis of LLR data (Williams et al., 2004) strongly limits such variations and constrains a local ($\sim$1 AU) scale expansion of the solar system as  $\dot a/a=-\dot G/G=(6\pm7) \times 10^{-13}$ yr$^{-1}$, including that due to cosmological effects. Interestingly, the achieved accuracy in $\dot G/G$ implies that, if this rate is representative of our cosmic history, then $G$ has changed by less than 1\% over the 13.7 Gyr age of the universe.  The ever-expanding LLR data set and its increasing accuracy will lead to further improvements in the search for spatial or temporal variability of G.  The next step in this direction is interplanetary ranging (see, e.g., Turyshev \& Williams, 2007) using laser transponders. Additionally, double pulsars might become competitive in measuring possible variations of $G$. Independent checks are, or course, important.

There is a connection between the variation of fundamental constants and a violation of the EP; in fact, the former almost always implies the latter. For example, should there be an ultra-light scalar particle, its existence would lead to variability of the fundamental constants, such as $\alpha$ and $m_e/m_p$. Because masses of nucleons are $\alpha$-dependent, by coupling to nucleons this particle would mediate an isotope-dependent long-range force \citep{Dvali-Zaldarriaga-2002,Uzan-2003}. The strength of this coupling is predicted to be within a few of orders of magnitude of existing experimental bounds for such forces; thus, the new force could potentially be measured in precision tests of the EP. Therefore, the existence of a new interaction mediated by a massless (or very low-mass) time-varying scalar field would lead to both the variation of fundamental constants and violation of the WEP, ultimately resulting in observable deviations from general relativity.

Following the arguments above, for macroscopic bodies, one expects that their masses depend on all the coupling constants of the four known fundamental interactions, which has consequences for the motion of a body. In particular, because the $\alpha$-dependence is {\it a priori} composition-dependent, any variation of the fundamental constants will entail a violation of the EP. This allows one to compare two classes of experiments---the clock-based $\dot \alpha$ experiment and the EP-testing one---in the search for variation of the parameter $\alpha$ in a model-independent way. In this comparison, the solar system EP tests emerged as the superior performers \citep{Nordtvedt-2002}.

\subsection{Do extra dimensions or other new physics alter the inverse square law?}

Many modern theories of gravity, including string, supersymmetry, and brane-world theories, suggest that new physical interactions will appear at short ranges.  This may happen, in particular, because at sub-millimeter distances new dimensions may exist, thereby changing the gravitational inverse-square law (ISL) \citep{Arkani-Hamed-etal-1999}. Similar forces that act at short distances are predicted in supersymmetric theories with weak scale compactifications \citep{Antoniadis-etal-1998}, in some theories with very low energy supersymmetry breaking \citep{Dimopoulos-Giudice-1996}, and also in theories of very low quantum gravity scale \citep{Sundrum-1999,Dvali-etal-2002}. These multiple predictions provide some of the motivation for experiments that would test for possible deviations from Newton's gravitational ISL at very short distances.

On {\it short distances}, recent torsion-balance experiments \citep{Kapner-etal-2007} tested the ISL at distances less than the dark-energy length scale  $\lambda_d=\sqrt[4]{\hbar c/u_d}\approx 85~\mu$m, with energy density $u_d\approx3.8$~keV/cm$^3$.  It was found that the ISL holds down to a length scale of 56~$\mu$m and that an extra dimension must have a size less than $44~\mu$m.  With these results experiments reached the level needed to test dark-energy physics in a laboratory; they also provided constraints on new forces \citep{Adelberger-etal-2007}, making such experiments very relevant and competitive with corresponding particle-physics research. 

Although most attention has focused on the behavior of gravity at short distances \citep{Adelberger-etal-2003}, tiny deviations from the ISL could well occur at much larger distances. The strong coupling phenomenon \citep{Dvali-2006} makes modified gravity theories predictive and potentially testable at scales that are much shorter than the current cosmological horizon.  Because of the key role that non-linearities play in some relativistic cosmologies, namely those of its scalar sector, their presence could lead to observable gravitational effects in our solar system.  In fact, there is a possibility that non-compact extra dimensions could produce observable deviations in the solar system \citep{Dvali-etal-2003}. 

By far the most stringent constraints on violation of the ISL on {\it large distances} to date come from very precise measurements of the lunar orbit about the Earth. Analysis of the LLR data tests the gravitational ISL to $3\times10^{-11}$ of the gravitational field strength on scales of the 385,000 km Earth-moon distance \citep{Muller-etal-2007}.  New LLR facilities \citep{Murphy-etal-2008} will reach sensitivity of $10^{-12}$, while millimeter-class laser ranging to Mars could extend the scale to 1.5 AU, while improving the corresponding ISL tests by more than three orders of magnitude. 

\subsection{What is the nature of spacetime?}

Given the immense challenge posed by the unexpected discovery of the accelerated expansion of the universe, it is important to explore every option to explain and probe the underlying physics.  Theoretical efforts in this area offer a rich spectrum of new ideas that can be tested by experiment. Motivated by the dark-energy and dark-matter problems, long-distance gravity modification is one of the radical proposals that has recently gained attention (Deffayet et al., 2002a).  Theories that modify gravity at cosmological distances exhibit a strong coupling phenomenon of extra graviton polarizations (Deffayet et al., 2002b, Dvali, 2006). The ``brane-induced gravity'' model (Dvali et al., 2000) provides an interesting way of modifying gravity at large distances to produce an accelerated expansion of the universe, without the need for a non-vanishing cosmological constant. One of the peculiarities of this model is the way one recovers the usual gravitational interaction at small (i.e. non-cosmological) distances, providing strong motivation for precision tests of gravity on solar system scales (Dvali et al., 2003). These ideas focus on the question: What is the nature of spacetime at various scales and how can it be tested? 

On {\it solar system scales}, the Eddington parameter, $\gamma$, whose value in general relativity is unity, is the most fundamental parameter in that $\frac{1}{2}(1-\gamma)$ is a measure, for example, of the fractional strength of the scalar gravity interaction in scalar-tensor theories of gravity (Damour \& Nordtvedt, 1993; Damour et al., 2002).  Specifically, the quantity $\frac{1}{2}(1-\gamma)$ defines corrections to the spacetime around massive bodies. Currently, the most precise value for this parameter, $\gamma-1=(2.1\pm2.3)\times10^{-5}$, was obtained using microwave tracking to the Cassini spacecraft (Bertotti et al., 2003) during a solar conjunction experiment. This accuracy approaches the region where multiple tensor-scalar gravity models, consistent with recent cosmological observations (Spergel et al., 2006), predict a lower bound for the present value of this parameter at the level of $(1-\gamma) \sim~ 10^{-6}-10^{-7}$.  Therefore, improving the measurement of $\gamma$ would provide crucial information to separate modern scalar-tensor theories of gravity from general relativity, probe possible ways for gravity quantization, and test modern theories of cosmological evolution (Turyshev, 2008).

Technology is available to enable strong gains in the measurement accuracy of the parameter $\gamma$ in solar system experiments. Thus, precision laser ranging between the Earth and a Mars lander performed with a transponder capable of reaching 1-mm precision could provide a measurement of $\gamma$ with accuracy of a part in $10^7$, limited by Martian dust. To reach accuracies beyond this level one needs special space experiments (see below). With sensitivities in measuring $\gamma-1$ down to $10^{-9}$, corresponding missions could significantly improve our knowledge of relativistic gravity, pushing to unprecedented accuracy the search for cosmologically relevant scalar-tensor theories of gravity by looking for a remnant scalar field in today's solar system (Turyshev, 2009). 

On {\it large scales}, studies of gravitational radiation can provide potentially powerful insight for large-distance modified gravity theories by directly observing the structure and dynamics of space and time. They will also be able to detect the signatures of cosmic superstrings or phase transitions in the early universe and contribute to the study of the dark energy that now dominates the evolution of the universe. Low-frequency gravitational wave detectors could detect and study a gravity wave signal from an in-spiraling massive binary that provides a ``standard siren'' allowing one to get a direct measure of a cosmic luminosity distance. The signals from black hole mergers will provide incisive tests of strong-field gravity. The possibility of observing stochastic primordial gravity waves is also very exciting; many other science results are expected. 

\section{Why is this decade important?}

The four key questions highlighted in this white paper are central to our understanding of Nature. If realized, the science opportunities presented here could significantly advance research in fundamental physics by either dramatically enhancing the range of validity of general relativity on various scales or leading to a spectacular discovery. Furthermore, present technological capabilities make the next decade especially ripe for progress in the area of gravitational physics.

Our possible return to the moon in the next decade would strongly advance the capabilities of LLR---potentially replacing the precision-limiting reflectors currently in place.  Recent successful interplanetary laser-link demonstrations \citep{Sun-etal-2005,Smith-etal-2006} show readiness of the new generation of active laser ranging for multitudes of science applications in the solar system \citep{Turyshev-Williams-2007}. On the moon, laser transponders would allow dozens of existing ground-based satellite laser ranging systems to contribute to this science. Placed on other bodies such as Mars, Mercury, or an asteroid, laser transponders would allow us to probe relativistic gravity to orders-of-magnitude higher precision (perhaps limited by planetary surface environments or/and asteroid ``noise''), especially were a primary laser transmitting system to be placed on the moon to avoid complications from the earth's atmosphere. 

Further tests of the WEP in free-fall from balloons, in suborbital flight, and in orbit offer the possibility of accuracy gains, listed in order of increasing accuracy (and also in order of technological readiness), from about two to five orders of magnitude. Long arm (low frequency) laser interferometric gravitational wave detectors and observations of some types of pulsar systems will offer insight into strong gravity regimes. Other missions with drag-free spacecraft performing interferometric laser measurements between spacecraft could improve measurements of $\gamma$ by four orders of magnitude over presently achieved accuracy.

Recent years have seen a remarkable maturation of a significant number of key technologies that would be useful in probes of gravitational physics. Among them are high-precision accelerometers and cryogenic differential displacement detectors, drag-free controls using He-proportional thrusters or small ion thrusters as actuators, ultra-stable lasers in space, He-dewars, cryo-coolers, superconducting detectors, short-pulse lasers, single-photon picosecond detector arrays, magnetic spectrometers, small trapped-ion clocks, lightweight H-maser clocks and atomic clocks using laser-cooled atoms. Some of these technologies have been space-qualified and some have already been flown in space supporting drag-free operations of the Gravity Probe B mission, thereby paving the way for new fundamental physics experiments. 

The state of technological readiness to tackle the science opportunities discussed in this white paper has changed markedly in the last decade.  Now is a good time to advance our exploration of various regimes of relativistic gravity.


\end{document}